# Prototyping Information Visualization in 3D City Models: a Model-based Approach

Claudine Métral[*], Gilles Falquet

Centre Universitaire d'informatique, University of Geneva,
7 route de Drize, 1227 Carouge, Switzerland
{claudine.metral gilles.falquet}@unige.ch

**Abstract.** When creating 3D city models, selecting relevant visualization techniques is a particularly difficult user interface design task. A first obstacle is that current geodata-oriented tools, e.g. ArcGIS, have limited 3D capabilities and limited sets of visualization techniques. Another important obstacle is the lack of unified description of information visualization techniques for 3D city models. If many techniques have been devised for different types of data or information (wind flows, air quality fields, historic or legal texts, etc.) they are generally described in articles, and not really formalized. In this paper we address the problem of visualizing information in (rich) 3D city models by presenting a model-based approach for the rapid prototyping of visualization techniques. We propose to represent visualization techniques as the composition of graph transformations. We show that these transformations can be specified with SPARQL construction operations over RDF graphs. These specifications can then be used in a prototype generator to produce 3D scenes that contain the 3D city model augmented with data represented using the desired technique.

**Keywords.** information visualization, 3D user interface design, 3D city models, model-based prototyping

## 1. Introduction

### 1.1 Motivation

3D city models are widely used all around the world. A current trend is adding semantics to the geometrical objects describing the city (buildings, transport elements, trees, city furniture, etc.) in order to enlarge the possible applications of such models. Adding semantics to 3D city models can be done according to various ways and for various additional information types. The added information can represent legal information, pollutant concentrations, noise levels, wind velocity, road names, historical facts, etc. A wide range of 3D visualization techniques is needed for visualizing such information ranging from text (legal information, historical facts, road names) to scalar fields (pollutant concentrations, noise levels) or flows (wind velocity). When creating 3D city models, selecting relevant information visualization techniques is far from trivial. This task should be eased by providing a prototyping tool enabling the designer to visualize, for such or such technique, his set of data in his 3D city model, and to conduct usability tests with different techniques.

---

[*] Corresponding author

The efficiency of the prototyping approach relies on the possibility to develop prototypes at a fraction of the cost of the final interface. This is precisely why this approach is difficult to apply in 3D UI: There is a lack of tools and techniques to develop prototypes at a low cost. For instance, the well known low-fidelity paper and pencil prototyping techniques cannot be adapted to the third dimension.

The question we address in this work is therefore: Can we develop models, algorithms, and tools to efficiently create prototypes for the visualization of data in 3D virtual environment? More precisely, is there a way to specify prototypes (specify the 3D model, data, and visualization technique) and then have a generator produce the actual prototype, instead of building it manually?

## 1.2 Overview

This paper starts with a brief state of the art on information visualization and on 3D user interface prototyping.

The next sections describe the approach that we have developed. We propose a model-based approach for the generation of prototypes for the visualization of information in 3D city models. The idea is to provide the 3D city model interface designer with (1) a modeling language to specify the visualization technique to be tested and (2) an automated generation tool that takes as input the technique specification, the reference 3D city model and a data set, and produces a 3D test environment where the given data set is visualized according to the specified technique. The central part of the technique specification model is intended to describe the mapping of data set elements to visual objects. Adapting the data state model of Chi (2000), this mapping is decomposed into several stages: extracting features from data; mapping the features to abstract visual objects and relations; mapping these objects and relations to concrete visual objects (layout generation). A layout manager, which is a constraint solver, produces the concrete visual objects and places them in the 3D city model (output rendering). In many cases this output rendering implies complex algorithms in order to compute the geometrical properties, shape, position... of the concrete visual objects.

We then conclude with a work planned in a near future: the implementation and classification of such algorithms for a representative set of 3D visualization techniques.

## 2. Background and related work

### 2.1 Information Visualization in 3D Environments

Visualizing semantically enriched 3D models consists in displaying the geometric part of the model (the geometry of the modeled spatial objects) and displaying visual objects that represent the enrichment information attached to these objects. There are two cases to distinguish, depending on the spatial or non-spatial nature of the enrichment information. Non-spatial information is attached to a 3D object but it doesn't have any spatial component. The maximum capacity of a room or the name of a street are non-spatial, they can be displayed at different places in the scene, provided the link with the described object is made obvious for the user. The visualization problem for this type of information is closely related to information and data visualization and geovisualization (Chen et al., 2008, Bleish 2012). The main problem here, which is not addressed in traditional information visualization techniques, is to find an adequate location to display the information. These questions have been explored, for instance, for the associations of labels to roads (Vaaraniemi et al, 2013) or the association of historical information to buildings (Alamouri and Pecchioli, 2010).

In the case of spatialized information, the information elements already have a precise spatial location, for instance the noise level measured at a given point on a facade, or the wind flow (a vector field) in a street canyon. The visualization of this type of information is generally called scientific visualization. It may rely on simple visualization techniques that represent the information (such as air quality values or number of pedestrians) by objects (e.g. solids) put at specific places in the 3D city model (San José et al, 2012) (Marina et al, 2012). It may also involve sophisticated algorithms to compute isosurfaces, flow lines, vortexes, etc. (Laramee et al, 2004), (Post et al, 2003) (Amorim et al, 2012). Many techniques have been developed to visualize this type of information and the main question is to select the most efficient one for a given 3D model and a given user task.

### 2.2 3D user interface prototyping

Research work related to the design of user interfaces (Shneiderman, 1998) or more specifically related to 3D user interfaces (Bowman, 2004) exist for a long time. 3D user interfaces (3D UI) can be the result of different development approaches: the programmatic approach where the 3D UI is obtained by coding, the toolkit-based approach where the developer directly codes and implements the final interface by using predefined sets of elements and objects provided by the toolkit, and the methodological approach that relies on existing numerous methods for developing 3D UI (Gonzalez Calleros, 2010). But these approaches are not adapted to the prototyping of information visualization in 3D city models.

# 3. A Model-based Approach for Prototyping 3D Information Visualization

## 3.1 Knowledge Sources

We studied the scientific literature about applications based on 3D city models to obtain a global view of the domain. Although the techniques used in these applications are often not explicitly described, they provide enough information to draw initial classification axis. The studied models and applications are used for various tasks, such as:

- Evaluation of the wind comfort for pedestrians in a city street (Amorim et al, 2012) where 3D coloured polylines (colour representing wind velocity) are added to the geometrical model.

- Assessment of air quality in a street or neighbourhood by adding coloured solid objects to the 3D buildings (Lu et al, 2009), (San José et al, 2012).

- Analysis of pedestrian behaviour (Marina et al, 2012) where colored bars visualize spatial distribution of pedestrian movement.

- Analysis of human perception of space (Fisher-Gewirtzman, 2012) where colored lines represent visual exposure or visual openness in the 3D city model.

- Visualization of historically enriched 3D city models where information (text and images) has been added to the geometrical model (Alamouri and Pecchioli, 2010), (Hervy et al, 2012).

From this study we can define different cases:

- Each data element is individually represented by an object (point, solid…) situated at the (x,y,z) location of the data, with parameters (color, size, transparency…) representing the value of the data (Lu et al, 2009), (San José et al, 2012), (Marina et al, 2012).

- A set of data is represented as a whole by a complex object such as a polyline (Amorim et al, 2012).

- Each data (eg a text) is individually represented by an object (panel of text…) associated to one or several objects (building…) of the 3D city model (Alamouri and Pecchioli, 2010), (Hervy et al, 2012).

- The data describe relations between different objects of the 3D city model and can be represented by objects (lines…) between these objects (windows…) (Fisher-Gewirtzman, 2012).

These cases can be classified according to (1) the type of data and (2) the type of representation:

- The dataset elements can be associated to a point or a region of space, to an object of the 3D city model or to a relation between several objects of the 3D city model.

- The dataset elements can be individually represented by a visual object such as a sphere, a panel…, either at their exact location or with an offset. Alternatively they can be globally represented: in this case a dataset is represented by a set of visual objects, such as isosurfaces for a scalar field.

**3.2 General Approach**

We selected as starting point an ontology of 3D visualization techniques (Métral et al, 2014) that provided of a common and formalized representation of visualization techniques in 3D environments. This ontology describes input data, output rendering and usage of visualization techniques. But it does not describe how the techniques map their input data to the output rendering. These mapping descriptions are precisely what we need to automatically generate the output rendering of a specific visualization technique. Figure 1 illustrates the main elements of such a mapping.

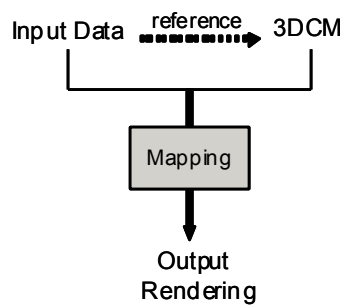

**Figure 1.** Visualisation technique as a mapping.

The question is thus, how to specify this kind of mapping?
   Our approach is based on the notion of graph transformation and on the distinction between abstract and concrete visual objects. In the following we will use of the resource description framework RDF (RDF, 2014) and the SPARQL query language (SPARQL, 2013) to present the approach and its concrete implementation. A 3D city model, input data elements and an output rendering are described as RDF graphs. A visualization is a graph transformation defined by (1) an abstract representation that is specified as a graph transformation expressed with SPARQL constructions and (2) a concrete representation generated with a layout manager. The abstract visualization produces abstract visual objects and relations. The layout manager makes the representation concrete by computing object attributes (e.g. their coordinates) that were left undefined at the abstract level.

A last step is the execution of an output script to generate the final visualization, and hence the prototype, in the desired 3D modeling language (X3D, COLLADA, KML, etc.). Figure 2 shows the whole generation process.

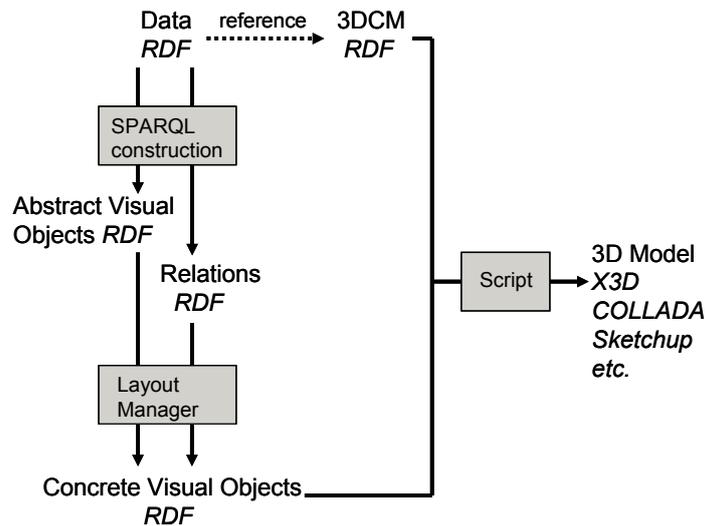

**Figure 2. Mapping description in detail**

## 4. RDF-based Specification of a Visualization Technique

We suppose that the 3D city models used for visualizing information are represented in CityGML (OGC 12-019, 2012). The CityGML XML Schema is almost compatible with RDF/XML, i.e. an XML file that conforms to the CityGML schema is almost an RDF/XML file. This comes from the fact that CityGML files obey to the alternating class-property structure: each element that represents a class entity (`bldg:Building`, `bldg:GroundSurface`, `gml:MultiSurface`, `gml:Polygon`, ...) contains only elements that represent properties (`bldg:boundedBy`, `bldg:lod2MultiSurface`, `gml:surfaceMember`, ...) and property elements don't have attributes. In addition, the `gml:id` attribute can serve as a resource identifier for the RDF (entities without `gml:id` will be mapped to RDF blank nodes). Therefore it is relatively straightforward to write a transformation script that generates an RDF version of a CityGML 3D city model.

## 4.1 Data Representation

As defined in (Métral et al, 2014) the information to visualize is composed of an information type (*PollutantConcentration*, *PedestrianCounting*, *IntervisibilityRate*...), a data type (*Integer, Vector, RichText*...) and an input location (*Point, Surface, CityObject*...). When these data represent measurement, their level of measurement (nominal, ordinal, quantitative) is in fact given by their datatype, more precisely, by the operations available (or not) for each datatype.

The input location – and thus the georeferencing of the data – is defined according to GML (OGC 07-036, 2007). The data elements are represented as RDF triples *subject property object*, depending of what these data represent. We distinguish three cases, depending on the spatial component of the considered data.

**Data elements associated to a point or to a region (spatial data)**

Each data element *x* is represented by a value *val* and by a location *loc*, which is a spatial object (not a city object) such as a point or a surface. The general RDF form of a spatial data element is:

    *x* a *DataElementClass*. *x* :value *val*. *x* :location *loc*.
    *loc* a *SpatialObjectClass* ; ... *properties of the location*
    *val* ... *properties of the value, for complex values*

Such data can be a pedestrian counting with *val* being the counting value and *loc* the area used for the counting, or a pollutant concentration with *val* being the pollutant concentration level and *loc* the point or region associated:

```
:data1 a :PollutantConcentration.
:data1 :value 5.67.
:data1 :location :loc1.
:loc1 a :Point; :xcord 4.5 ; :ycoord 44 ; zcoord 1.5 .
```

**Data elements related to an object of the 3D city model**

Each data element *x* is represented by a value *val* and is related to the city object (*Building*, *Window*, *Vegetation*…) to which it refers. The general schema is

    *x* a *DataElementClass*. *x* :value *val*. *x* :about *y*.
    *y* a *CityObjectClass* ; ... *properties of the city object*

Such data can be the description of a building with *val* being the description text and *y* the related building:

```
:data2 a :RichText.  :data2 :value "....".
:data2 :about :co2.
:co2 a cgml:Building ; gml:geometry ...
```

In general the city object will be described in a separate RDF graph that represent the 3D city model.

**Data elements representing a relation between objects of the 3D city model**

Each data element *x* is represented by an optional value *val* and properties (`arg1`, `arg2`, ...) that point to the city objects (*Building*, *Window*, *Vegetation*...) it relates:

> *x* a *RelationType*.   *x* :value val.   *x* :arg1 $y_1$.   *x* :about $y_2$.   $y_1$ a *CityObjectClass$_1$*.   $y_2$ a *CityObjectClass$_2$*. ...

Such data can represent the intervisibility between windows with *val* being the intervisibility level, and $y_1$ and $y_2$ the related windows:

```
:data3 a :IntervisibilityRelation.   :data3 :value 0.3 .
:data3 :arg1 :win1 ; :arg2 win2.
:win1  a cgml:Window ; ... . :win2 a cgml:Window ; ... .
```

### 4.2 Abstract Level

The abstract level mapping associates abstract visual objects and relations to input data (see Figure 2). The abstract vocabulary comprises simple visual objects (*Sphere*, *Line*, ...), complex objects (*IsoSurfaces*, *FlowLines*, ...), and abstract spatial relations (*near*, *above*, *frontOf*, *inside*...). Each type of abstract objects has specific visual properties (color, shape, center, ...) that can be set a this level or left undefined. The abstract vocabulary may be arbitrarily extended with new elements, provided they are taken into account at the concrete level.

Since we represent data and the city model as RDF graphs, it is natural to use the SPARQL language to specify the graph transformation that corresponds to the abstract level mapping. The SPARQL query language provides several query forms (SPARQL, 2013). The *select* query form returns all (or a subset of) the variables bound in a query pattern match. The *construct* query form returns an RDF graph. The graph is built by substituting variables in a set of triple templates. A query consists in two parts: the *select* or *construct* clause and a *where* clause that provides the basic graph pattern to match against the data graph. In the SPARQL queries below all elements of the RDF graph that satisfy the *where* statement are selected. Then, for each of them, a new RDF graph is generated with the *construct* statement. We thus obtain a blank node of a certain abstract visual object type and with some visual properties. In fact, in the *select* statement we have the same RDF graph as the one used for the description of the input data, but with *loc*, *val*... as variables (this is why they are noted *?val*, *?loc*...).

**Data elements associated to a point or to a region: individual representation**

In this case the blank node (named `_:1`) is associated to the same location *loc* as the data element. Its visual properties are derived from the data value ($f_1, f_2, ...$ represent computations on *?val* and its properties).

```
construct {_:1 a AbstractVisualObjectType .
    _:1 visualProperty₁ f₁(?val). _:1 visualProperty₂ f₂(?val). ...
```

```
  _:1 :location ?loc}
from DataGraph
where {?x a DataElementClass.  ?x :value ?val.  ?x :location
?loc}
```

Such data elements can be a pedestrian counting represented by a cone or a pollutant concentration (with the value *val* being the pollutant concentration level and location *loc* being the point associated) represented by a sphere located at *loc* and whose radius is computed from *val*:

```
construct {_:1 a :Sphere .
  _:1 :radius ?val/100.
  _:1 :location ?loc}
from DataGraph
where {?x a :PollutantConcentration.  ?x :value ?val.
  ?x :location ?loc.  ?loc a :Point}
```

**Data elements associated to a point or to a region: global representation**
In this case of a global representation a (complex) visual object represents the whole dataset:

```
construct {:vObj a AbstractComplexVisualObjectType .
:vObj :inputData _:1.  _:1 :value ?val.  _:1 :location ?loc}
from DataGraph
where {?x a DataElementClass.  ?x :value ?val.  ?x :location
?loc}
```

Each data element becomes an input data for the complex visual object :vObj, keeping its original location and value. Typical abstract objects of this type are isosurfaces, flowlines, stream ribbons, etc.

**Data elements related to an object of the 3D city model**
In the case of data elements related to a city object the location is in fact a blank node (named _:2) defined by a spatial relation (*near, above, inside…*) and by the related city object:

```
construct {_:1 a AbstractVisualObjectType .
  _:1 visualProperty₁ f₁(?val).   _:1 visualProperty₂ f₂(?val).   …
  _:1 :location _:2.
  _:2 a SpatialRelation; :arg1 _:1; :arg2 ?y }
from DataGraph
where {?x a DataElementClass.  ?x :value ?val.  ?x :about ?y.
?y a CityObjectClass}
```

For a text that has to be displayed on a panel located near the building it describes we can define:

```
construct {_:1 a :Panel .
  _:1 :content ?val.
  _:1 :location _:2.
  _:2 a :nearRealtion; :arg1 ?y; :arg2 _:1 . }
from DataGraph
```

```
where {?x a :RichText.   ?x :value ?val.   ?x :about y.   ?y a
cgml:Building}
```

**Data elements representing a relation between objects of the 3D city model**
In the case of data elements representing a relation between objects of the 3D city model, what we have to represent is a connection between the city objects (represented by a blank node _:2 linking the different elements together):

```
construct {_:1 a AbstractVisualObjectType .
_:1 visualProperty₁ f₁(?val, ?y1, ?y2).
_:1 visualProperty₂ f₂(?val, ?y1, ?y2).
...
from DataGraph
where {?x a DataElement.   ?x :value ?val.
       ?x :arg1 ?y1.    ?x :arg2 ?y2.
       ?y1 a CityObject.   ?y2 a CityObject. ...}
```

An example is the representation of intervisibility levels between windows by colored lines drawn between the related windows:

```
construct {_:1 a :Line .
_:1 :color [a :Color; :red ?val; :green 0; blue 0].
_:1 :endpoint ?y1.  _:1 :endpoint ?y2.
from DataGraph
where {?x a :IntervisibilityRelation.   ?x :value ?val.
       ?x :arg1 ?y1.   ?x :arg2 ?y2.
       ?y1 a cgml:Window.   ?y2 a cgml:Window}
```

Note that this representation is abstract because the line endpoints are objets (windows) not precise points in the 3D space.

### 4.3 Concrete level

The concrete level associates abstract visual objects to input data using the abstract visual objects and the relations generated at the abstract level (see Figure 2).

**Data elements associated to a point or to a region: individual representation**
In the case of a data element directly associated to a point it is possible to define a SPARQL statement similar to the one defined at the abstract level. In the case of a data element associated to a surface we have first to compute a point inside the surface and then to use this point for precisely positioning the concrete visual object.

```
construct {_:1 a ConcreteVisualObjectType .
_:1 visualProperty₁ f₁(?val).   _:1 visualProperty₂ f₂(?val).  ... }
_:1 preciseLocationf(?loc) .
from DataGraph
where {?x a DataElementClass.   ?x :value ?val.   ?x :location
?loc}
```

**Data elements associated to a point or to a region: global representation**
In the case of such a global representation we have to compute one or several visual objects that can have several parameters depending not only on the data elements but also on their abstract representation, and that will be placed at a precise location in the 3D scene. For example an abstract isosurface can be represented by several concrete multisurfaces whose number depends on the value intervals of the isosurface.

**Data elements related to an object of the 3D city model**
In the case of a data element associated to a city object we have to compute a point that will be used for precisely positioning the concrete visual object in the scene. This computation has to take in account the city object and the spatial relation between it and the abstract visual object. Some parameters have also to be computed. For the example of a text panel that has to be placed near a building the computed point can be used for positioning the panel. Parameters such as height or orientation of the panel have also to be computed.

**Data elements representing a relation between objects of the 3D city model**
In the case of a data element representing a semantic relation between several city objects we have to compute a point in each of these city objects. These points will be used for precisely positioning the concrete visual object that will rely them. Some parameters may also be computed. For the example of intervisibility lines between windows the computed points can be the center of the windows in order to draw a line between these points while parameters can define a color based on the intervisibility rate.

As we have seen with previous examples, using SPARQL queries for generating the concrete visual objects is only possible in very simple cases. For most cases we need a layout manager whose task is to compute precise locations as well as the needed parameters.

## 5. Layout manager

The layout manager is a constraint solver, that produces concrete visualization objects (in X3D, COLLADA, SketchUp...). The layout manager must conform to the abstraction principle, i.e. it must always be possible to reconstruct the abstract view from the concrete one. The layout managers tasks for the different cases are described below.

**Data elements associated to a point or to a region: individual representation**
The main task for such data elements is to transform data coordinates into output coordinates

**Data elements associated to a point or to a region: global representation**
The layout managers for these objects are sophisticated algorithms that compute surfaces for representing scalar fields, or lines for representing trajectories in vector fields, etc.

**Data elements related to an object of the 3D city model**
The layout manager must solve placement relations, i.e. find a location in the output space that clearly represents an abstract spatial relation. For example the *above* relation between a panel and a building can be represented by placing the bottom-centre of the panel 2 meters higher than the highest point of the building roof surfaces and at the barycentre of the building ground surface.

**Data elements representing a relation between objects of the 3D city model**
For semantic relations between city objects the layout manager must find locations in the city objects involved. For example, if the intervisibility between two windows is to be represented by a straight line segment, the layout manager must select appropriate points in each window surface as endpoints for the line.

## 6. Implementation

### 6.1 Generation process

The generation process proceeds as follows:

1. Transform the 3D city model into an RDF graph and upload it in a triple store able to store and manage RDF data.

2. Transform each dataset into an RDF graph and upload it in the triple store (as a separate graph).

3. For each dataset (RDF graph) apply the desired visualization transformation to produce a graph of abstract visual entities (objects and relations). To some extent this can be done dirctly with SPARQL queries. Indeed, in the current version of SPARQL, the *construct* query form doesn't support operations (such as *?val/100*) or functions (such as *concat*). To overcome this limitation some scripting or computing is needed (we can use the *select* query form that allows such operations with some scripting since the *select* query doesn't directly generate a RDF graph).

4. Apply a layout manager to compute the concrete representation (e.g. X3D or SketchUp objects) of each abstract visual entity.

Uploading datasets to the RDF store may require a "dictionary" to map the city object references present in a dataset to the corresponding nodes of the 3D city model graph (identified by their `gml:id`).

**6.2 Specifying and Implementing the layout manager**

**For abstract objects located in a spatial region**
If the region is a point the mapping is immediate, there is no real choice as to where the concrete object must be. If the region is a point set (line, surface, volume) the layout manager must choose a point within the region to place the concrete object. Thus the specification of such a layout manager amounts to selecting a function that maps a spatial region to a point (or more generaly to a region)

**For abstract objects linked to a city object through a spatial relation**
The layout manager is essentially a mapping from a city object and a relation to a point that represents and satisfies this relation. For instance, the *above* relation for a building should yield a point that is higher than the building and above its ground surface. Of course each spatial relation may be represented by several different mappings, so the specification of a layout manager consists in choosing one such mapping (e.g. in a library of mappings)

**For abstract objects representing relations between city objects**
The choice function will generally rely on geometric functions such as min, max, barycenter, etc. A layout manager is in charge of mapping an (set of) abstract visual object to a concrete user interface object.

The mapping computed by the layout manager aims to produce precise locations in the 3D city model. Alternatively the mapping could generate a concrete object with some parameters left undefined. For instance, a sphere whose radius is fixed but not its center point, a panel whose size is fixed but not its base point. The user thus could have the possibility to place the concrete object at the more suitable place according to his criteria.

## 7. Testing the Approach

To test the proposed model-based approach we selected a city model expressed in CityGML. This model represents a part of the city of Carouge in Switzerland and is composed of about one hundred buildings at LOD 2. Only buildings are represented in the model. We have used the Sesame triplestore for the storage and the retrieval of the RDF triples. Once loaded on the Sesame platform our model consists of about 80000 RDF triples.

A test case consists in using the approach to 1) specify a data set and a visualization technique (to check the expressiveness of the specification method), 2) to generate the test scene (to check if it correctly represents the visualization technique)

We present here a basic test case related to pedestrian counting. The data set contains the numbers of pedestrians counted at different locations. The visualization is made of cones whose location and height represent the counting location and value respectively. The output rendering is made of X3DOM elements corresponding to the cones and the city model geometry.

**Data representation**

The location of a data element is a 3D point with coordinates `xloc`, `yloc`, and `zloc`. The value is an integer number. The RDF representation of 42 pedestrians counted at $x = -13, y = 25, z = 0$ is

```
:pednum1  a :PedestrianCounting .
:pednum1 :value 42 .
:pednum1 :location :loc1 .
:loc1 :xloc -13.  :loc1 :yloc 25. :loc1 :zloc 0 .
```

**Abstract level**

The abstract representation is made of objects of type `Cone` with height and location directly drawn from the data. Here is the SPARQL query that generates these abstract objects (as two blank nodes named `_:1` and `_:2`):

```
construct {_:1 a :Cone.
 _:1 :height ?val.
 _:1 :location _:2. _:2 :xloc ?x. _:2 :yloc ?y. _:2 :zloc ?z}
where {?x a :PedestrianCounting.
       ?x :value ?val.
       ?x :location ?loc.
       ?loc :xloc ?x.  ?loc :yloc ?y. ?loc :zloc ?z}
```

Applying this query on the above data sample will generate

```
_:1 a :Cone.
_:1 :height 42.
_:1 :location _:2. _:2 :xloc -13.  _:2 :yloc 25. _:2 :zloc 0}
```

**Concrete level**

Even with the exact coodinates known we have to do some scripting since we have to transform the *x, y, z* numeric coordinates into the string "*x y z+val/2*" and the integer *height* value into the string "*height*" in order to generate correct X3DOM code. Here is the XML code with the values inset into the strings:

```
<transform rotation="1 0 0 1.5708" translation="-13 25 21">
   <shape>
      <appearance>
         <material diffusecolor="0 0 1"></material>
      </appearance>
      <cone height="42"></cone>
   </shape>
</transform>
```

Figure 3 illustrates the previous data displayed as a cone in the 3D city model (and dispayed with another pedestrian counting data) :

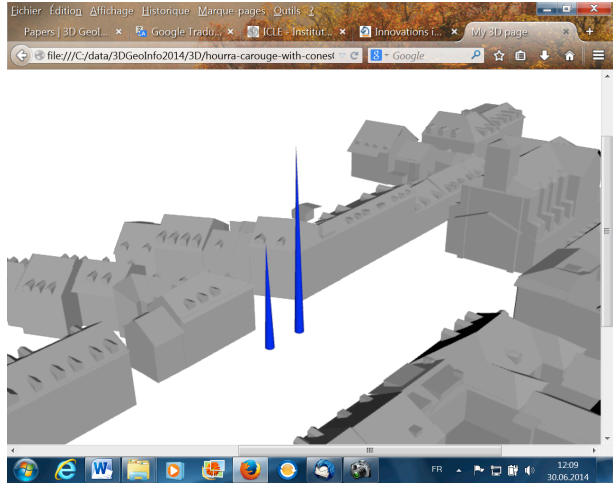

**Figure 3.** Visualizing pedestrian counts as cones

## Conclusion

In this paper we addressed the development of (rapid) prototyping tools for information visualization in 3D city models. A prototyping tool enables a designer to visualize in a specific 3D city model a specific set of data according to a specific visualization technique, and to conduct usability tests with different techniques.

We proposed a mapping of data set elements to visual objects in two stages: (1) a mapping of data features to abstract visual objects and relations, and (2) a mapping of these objects and relations to concrete visual objects (output rendering). A constraint solver named layout manager produces the concrete visual objects and places them in the 3D city model. In many cases the output rendering implies complex algorithms since the concrete visual objects are not simple 3D shapes but visual objects that embed functions to compute their geometrical properties, shape, position, etc. In a near future we plan to implement and classify such algorithms for a representative set of 3D visualization techniques.

## Acknowledgements

The model used for illustrating some visualization techniques relates to Carouge city (Switzerland) and has been provided by the Service de la mensuration officielle de l'Etat de Genève (SEMO).